\documentclass{aastex63}
\usepackage{graphicx}
\usepackage{natbib}
\usepackage{apjfonts}
\usepackage{amsmath}
\usepackage{color}
\usepackage{comment}
\usepackage{rotating}
\bibpunct{(}{)}{;}{a}{}{,}
\usepackage{lineno}
\usepackage[normalem]{ulem}

\reportnum{DES-2020-0541}
\reportnum{FERMILAB-FN-1101-AE}

\begin{document}
\title{The Diffuse Light Envelope of Luminous Red Galaxies} 
%\author{ Y. L. Leung$^1$, Y. Zhang$^{2, 1}$ et al. }
\vspace{1cm} (DES Collaboration) \\
% Author list file generated with: mkauthlist 1.2.3 
% mkauthlist -j apj -s lrg_diffuse.csv lrg_diffuse.tex 
\author{Y.~Leung}
\affiliation{Kavli Institute for Cosmological Physics, University of Chicago, Chicago, IL 60637, USA}
\author{Y.~Zhang}
\affiliation{Fermi National Accelerator Laboratory, P. O. Box 500, Batavia, IL 60510, USA}
\author{B.~Yanny}
\affiliation{Fermi National Accelerator Laboratory, P. O. Box 500, Batavia, IL 60510, USA}
\author{K.~Herner}
\affiliation{Fermi National Accelerator Laboratory, P. O. Box 500, Batavia, IL 60510, USA}
\author{J.~Annis}
\affiliation{Fermi National Accelerator Laboratory, P. O. Box 500, Batavia, IL 60510, USA}
\author{A.~Palmese}
\affiliation{Fermi National Accelerator Laboratory, P. O. Box 500, Batavia, IL 60510, USA}
\affiliation{Kavli Institute for Cosmological Physics, University of Chicago, Chicago, IL 60637, USA}
\author{H.~Sampaio-Santos}
\affiliation{Observat\'orio Nacional, Rua Gal. Jos\'e Cristino 77, Rio de Janeiro, RJ - 20921-400, Brazil}
\affiliation{Laborat\'orio Interinstitucional de e-Astronomia - LIneA, Rua Gal. Jos\'e Cristino 77, Rio de Janeiro, RJ - 20921-400, Brazil}
\author{V.~Strazzullo}
\affiliation{Faculty of Physics, Ludwig-Maximilians-Universit\"at, Scheinerstr. 1, 81679 Munich, Germany}
\author{M.~Aguena}
\affiliation{Laborat\'orio Interinstitucional de e-Astronomia - LIneA, Rua Gal. Jos\'e Cristino 77, Rio de Janeiro, RJ - 20921-400, Brazil}
\affiliation{Departamento de F\'isica Matem\'atica, Instituto de F\'isica, Universidade de S\~ao Paulo, CP 66318, S\~ao Paulo, SP, 05314-970, Brazil}
\author{S.~Allam}
\affiliation{Fermi National Accelerator Laboratory, P. O. Box 500, Batavia, IL 60510, USA}
\author{S.~Avila}
\affiliation{Instituto de Fisica Teorica UAM/CSIC, Universidad Autonoma de Madrid, 28049 Madrid, Spain}
\author{E.~Bertin}
\affiliation{Sorbonne Universit\'es, UPMC Univ Paris 06, UMR 7095, Institut d'Astrophysique de Paris, F-75014, Paris, France}
\affiliation{CNRS, UMR 7095, Institut d'Astrophysique de Paris, F-75014, Paris, France}
\author{S.~Bhargava}
\affiliation{Department of Physics and Astronomy, Pevensey Building, University of Sussex, Brighton, BN1 9QH, UK}
\author{D.~Brooks}
\affiliation{Department of Physics \& Astronomy, University College London, Gower Street, London, WC1E 6BT, UK}
\author{D.~L.~Burke}
\affiliation{SLAC National Accelerator Laboratory, Menlo Park, CA 94025, USA}
\affiliation{Kavli Institute for Particle Astrophysics \& Cosmology, P. O. Box 2450, Stanford University, Stanford, CA 94305, USA}
\author{A.~Carnero~Rosell}
\affiliation{Centro de Investigaciones Energ\'eticas, Medioambientales y Tecnol\'ogicas (CIEMAT), Madrid, Spain}
\affiliation{Laborat\'orio Interinstitucional de e-Astronomia - LIneA, Rua Gal. Jos\'e Cristino 77, Rio de Janeiro, RJ - 20921-400, Brazil}
\author{M.~Carrasco~Kind}
\affiliation{National Center for Supercomputing Applications, 1205 West Clark St., Urbana, IL 61801, USA}
\affiliation{Department of Astronomy, University of Illinois at Urbana-Champaign, 1002 W. Green Street, Urbana, IL 61801, USA}
\author{J.~Carretero}
\affiliation{Institut de F\'{\i}sica d'Altes Energies (IFAE), The Barcelona Institute of Science and Technology, Campus UAB, 08193 Bellaterra (Barcelona) Spain}
\author{M.~Costanzi}
\affiliation{INAF-Osservatorio Astronomico di Trieste, via G. B. Tiepolo 11, I-34143 Trieste, Italy}
\affiliation{Institute for Fundamental Physics of the Universe, Via Beirut 2, 34014 Trieste, Italy}
\author{L.~N.~da Costa}
\affiliation{Observat\'orio Nacional, Rua Gal. Jos\'e Cristino 77, Rio de Janeiro, RJ - 20921-400, Brazil}
\affiliation{Laborat\'orio Interinstitucional de e-Astronomia - LIneA, Rua Gal. Jos\'e Cristino 77, Rio de Janeiro, RJ - 20921-400, Brazil}
\author{S.~Desai}
\affiliation{Department of Physics, IIT Hyderabad, Kandi, Telangana 502285, India}
\author{H.~T.~Diehl}
\affiliation{Fermi National Accelerator Laboratory, P. O. Box 500, Batavia, IL 60510, USA}
\author{P.~Doel}
\affiliation{Department of Physics \& Astronomy, University College London, Gower Street, London, WC1E 6BT, UK}
\author{T.~F.~Eifler}
\affiliation{Jet Propulsion Laboratory, California Institute of Technology, 4800 Oak Grove Dr., Pasadena, CA 91109, USA}
\affiliation{Department of Astronomy/Steward Observatory, University of Arizona, 933 North Cherry Avenue, Tucson, AZ 85721-0065, USA}
\author{S.~Everett}
\affiliation{Santa Cruz Institute for Particle Physics, Santa Cruz, CA 95064, USA}
\author{B.~Flaugher}
\affiliation{Fermi National Accelerator Laboratory, P. O. Box 500, Batavia, IL 60510, USA}
\author{J.~Frieman}
\affiliation{Kavli Institute for Cosmological Physics, University of Chicago, Chicago, IL 60637, USA}
\affiliation{Fermi National Accelerator Laboratory, P. O. Box 500, Batavia, IL 60510, USA}
\author{J.~Garc\'ia-Bellido}
\affiliation{Instituto de Fisica Teorica UAM/CSIC, Universidad Autonoma de Madrid, 28049 Madrid, Spain}
\author{E.~Gaztanaga}
\affiliation{Institut d'Estudis Espacials de Catalunya (IEEC), 08034 Barcelona, Spain}
\affiliation{Institute of Space Sciences (ICE, CSIC),  Campus UAB, Carrer de Can Magrans, s/n,  08193 Barcelona, Spain}
\author{D.~Gruen}
\affiliation{SLAC National Accelerator Laboratory, Menlo Park, CA 94025, USA}
\affiliation{Kavli Institute for Particle Astrophysics \& Cosmology, P. O. Box 2450, Stanford University, Stanford, CA 94305, USA}
\affiliation{Department of Physics, Stanford University, 382 Via Pueblo Mall, Stanford, CA 94305, USA}
\author{R.~A.~Gruendl}
\affiliation{Department of Astronomy, University of Illinois at Urbana-Champaign, 1002 W. Green Street, Urbana, IL 61801, USA}
\affiliation{National Center for Supercomputing Applications, 1205 West Clark St., Urbana, IL 61801, USA}
\author{J.~Gschwend}
\affiliation{Laborat\'orio Interinstitucional de e-Astronomia - LIneA, Rua Gal. Jos\'e Cristino 77, Rio de Janeiro, RJ - 20921-400, Brazil}
\affiliation{Observat\'orio Nacional, Rua Gal. Jos\'e Cristino 77, Rio de Janeiro, RJ - 20921-400, Brazil}
\author{G.~Gutierrez}
\affiliation{Fermi National Accelerator Laboratory, P. O. Box 500, Batavia, IL 60510, USA}
\author{K.~Honscheid}
\affiliation{Center for Cosmology and Astro-Particle Physics, The Ohio State University, Columbus, OH 43210, USA}
\affiliation{Department of Physics, The Ohio State University, Columbus, OH 43210, USA}
\author{D.~J.~James}
\affiliation{Center for Astrophysics $\vert$ Harvard \& Smithsonian, 60 Garden Street, Cambridge, MA 02138, USA}
\author{A.~G.~Kim}
\affiliation{Physics Division, Lawrence Berkeley National Laboratory, Berkeley, CA 94720, USA}
\author{K.~Kuehn}
\affiliation{Australian Astronomical Optics, Macquarie University, North Ryde, NSW 2113, Australia}
\affiliation{Lowell Observatory, 1400 Mars Hill Rd, Flagstaff, AZ 86001, USA}
\author{N.~Kuropatkin}
\affiliation{Fermi National Accelerator Laboratory, P. O. Box 500, Batavia, IL 60510, USA}
\author{M.~Lima}
\affiliation{Departamento de F\'isica Matem\'atica, Instituto de F\'isica, Universidade de S\~ao Paulo, CP 66318, S\~ao Paulo, SP, 05314-970, Brazil}
\affiliation{Laborat\'orio Interinstitucional de e-Astronomia - LIneA, Rua Gal. Jos\'e Cristino 77, Rio de Janeiro, RJ - 20921-400, Brazil}
\author{M.~A.~G.~Maia}
\affiliation{Laborat\'orio Interinstitucional de e-Astronomia - LIneA, Rua Gal. Jos\'e Cristino 77, Rio de Janeiro, RJ - 20921-400, Brazil}
\affiliation{Observat\'orio Nacional, Rua Gal. Jos\'e Cristino 77, Rio de Janeiro, RJ - 20921-400, Brazil}
\author{R.~Miquel}
\affiliation{Instituci\'o Catalana de Recerca i Estudis Avan\c{c}ats, E-08010 Barcelona, Spain}
\affiliation{Institut de F\'{\i}sica d'Altes Energies (IFAE), The Barcelona Institute of Science and Technology, Campus UAB, 08193 Bellaterra (Barcelona) Spain}
\author{R.~L.~C.~Ogando}
\affiliation{Laborat\'orio Interinstitucional de e-Astronomia - LIneA, Rua Gal. Jos\'e Cristino 77, Rio de Janeiro, RJ - 20921-400, Brazil}
\affiliation{Observat\'orio Nacional, Rua Gal. Jos\'e Cristino 77, Rio de Janeiro, RJ - 20921-400, Brazil}
\author{F.~Paz-Chinch\'{o}n}
\affiliation{National Center for Supercomputing Applications, 1205 West Clark St., Urbana, IL 61801, USA}
\affiliation{Institute of Astronomy, University of Cambridge, Madingley Road, Cambridge CB3 0HA, UK}
\author{A.~A.~Plazas}
\affiliation{Department of Astrophysical Sciences, Princeton University, Peyton Hall, Princeton, NJ 08544, USA}
\author{A.~K.~Romer}
\affiliation{Department of Physics and Astronomy, Pevensey Building, University of Sussex, Brighton, BN1 9QH, UK}
\author{A.~Roodman}
\affiliation{SLAC National Accelerator Laboratory, Menlo Park, CA 94025, USA}
\affiliation{Kavli Institute for Particle Astrophysics \& Cosmology, P. O. Box 2450, Stanford University, Stanford, CA 94305, USA}
\author{E.~S.~Rykoff}
\affiliation{Kavli Institute for Particle Astrophysics \& Cosmology, P. O. Box 2450, Stanford University, Stanford, CA 94305, USA}
\affiliation{SLAC National Accelerator Laboratory, Menlo Park, CA 94025, USA}
\author{E.~Sanchez}
\affiliation{Centro de Investigaciones Energ\'eticas, Medioambientales y Tecnol\'ogicas (CIEMAT), Madrid, Spain}
\author{V.~Scarpine}
\affiliation{Fermi National Accelerator Laboratory, P. O. Box 500, Batavia, IL 60510, USA}
\author{M.~Schubnell}
\affiliation{Department of Physics, University of Michigan, Ann Arbor, MI 48109, USA}
\author{S.~Serrano}
\affiliation{Institut d'Estudis Espacials de Catalunya (IEEC), 08034 Barcelona, Spain}
\affiliation{Institute of Space Sciences (ICE, CSIC),  Campus UAB, Carrer de Can Magrans, s/n,  08193 Barcelona, Spain}
\author{I.~Sevilla-Noarbe}
\affiliation{Centro de Investigaciones Energ\'eticas, Medioambientales y Tecnol\'ogicas (CIEMAT), Madrid, Spain}
\author{M.~Smith}
\affiliation{School of Physics and Astronomy, University of Southampton,  Southampton, SO17 1BJ, UK}
\author{E.~Suchyta}
\affiliation{Computer Science and Mathematics Division, Oak Ridge National Laboratory, Oak Ridge, TN 37831}
\author{M.~E.~C.~Swanson}
\affiliation{National Center for Supercomputing Applications, 1205 West Clark St., Urbana, IL 61801, USA}
\author{T.~N.~Varga}
\affiliation{Max Planck Institute for Extraterrestrial Physics, Giessenbachstrasse, 85748 Garching, Germany}
\affiliation{Universit\"ats-Sternwarte, Fakult\"at f\"ur Physik, Ludwig-Maximilians Universit\"at M\"unchen, Scheinerstr. 1, 81679 M\"unchen, Germany}

\begin{abstract}
    We use a stacking method to study the radial light profiles of luminous red galaxies (LRGs) at redshift $\sim 0.62$ and $\sim 0.25$, out to a radial range of 200 kpc. We do not find noticeable evolution of the profiles at the two redshifts. The LRG profiles appear to be well approximated by a single Sersic profile, although some excess light can be seen outside 60 kpc. We quantify the excess light by measuring the integrated flux and find that the excess is about 10\% -- a non-dominant but still nonnegligible component.
\end{abstract}

\section{Introduction}

Studies have found evidence that there exists an envelope distribution of diffuse  stars around both early and late type galaxies \citep[e.g.,][]{ 1994Natur.370..441S, 2004MNRAS.347..556Z, 2007ApJ...671.1591I, article, 2013MNRAS.431.1121T, 2014MNRAS.443.1433D, 2015MNRAS.446..120D}, and an excessive amount around the central galaxies of galaxy clusters \citep[e.g.,][]{1986ApJS...60..603S, 2005MNRAS.358..949Z, 2015MNRAS.449.2353B, 2019ApJ...874..165Z}. These diffuse stellar envelopes provide clues to how frequently galaxies interact, and how the galaxy interactions affect the galaxy stellar distribution \citep{2011MNRAS.416.2802F, 2013MNRAS.434.3348C, 2018MNRAS.479.4004E}.  Unfortunately, their faintness and the effect of the observational point-spread function (PSF) mean that it is often difficult to quantify these diffuse stellar distributions accurately \citep{2008MNRAS.388.1521D,2014A&A...567A..97S, 2015A&A...577A.106S, 2012MNRAS.421..190Z}, and their detection sparked debates. 

We assess the diffuse stellar extent of luminous red galaxies (LRGs) and examine its redshift evolution between redshift $\sim$0.25 and $\sim$0.625. We use a stacking method with optical images \citep{2019ApJ...874..165Z} from the Dark Energy Survey (DES) to acquire high signal-to-noise LRG {\bf S}urface {\bf B}rightness (SB) measurements out to the radius of 200 kpc. We do not notice significant differences in the LRG profiles between redshift $\sim0.25$ and $\sim0.625$. This analysis assumes a $\Lambda$CDM cosmology model with the Hubble parameter $h=0.7$ and the matter density parameter $\Omega_m = 0.3$.

\section{Data and Methods}

The LRG sample used in the analysis is selected from DES Year 1 data by the redMagic algorithm \citep{2016MNRAS.461.1431R}, which is based on comparing galaxy colors to spectroscopic LRG samples. The algorithm delivers excellent galaxy photometric redshift estimation with a median scatter of $0.017(1+z)$, and provides the fiducial sample for the DES galaxy clustering analysis  \citep{2018PhRvD..98d2006E}.
We study the LRGs in a low redshift range with photometric redshifts between 0.24 and 0.26 ($z\sim 0.25$), and in a high redshift range between 0.62 and 0.63 ($z\sim 0.625$). We analyze the $r$ band SB profile of the $z\sim 0.25$ LRG sample, and the $i$ profile of the $z\sim 0.625$ sample. The red-shifting from $\sim 0.25$ to $\sim 0.625$ places the DES $r$ band at a similar rest-frame wavelength range with the $i$ band, and thus eliminates the need of performing K-corrections when examining redshift evolution.
 
\begin{figure*}
\includegraphics[width=1\linewidth]{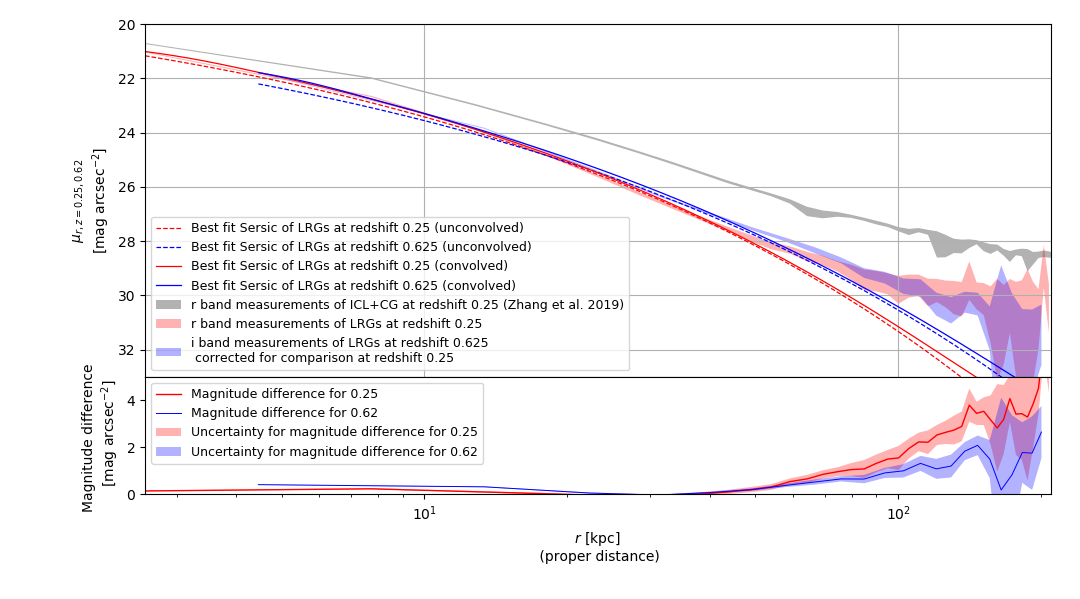}
\caption{The upper panel shows the measured LRG SB profiles at redshift $\sim$0.25 and $\sim$0.625. The blue and red shaded regions indicate the corresponding uncertainties. The solid and dotted lines show the same best-fit Sersic models with and without including the effect of PSF. The lower panel shows the residual between the data and the best-fit models. For comparison, we also overplot the intra-cluster light and central galaxy (ICL+CG) measurements at $z\sim 0.25$ in \cite{2019ApJ...874..165Z} as the grey shaded region.}
\label{fig:profile}
\end{figure*}

Our methods of preparing LRG images from DES Year1 to 3 observations and measuring their SB profiles closely follow the procedures in \cite{2019ApJ...874..165Z}. In total, we stack 201 LRGs at $z\sim 0.25$, and 1381 LRGs at $z\sim 0.625$. To reduce noise in the measurements, we also eliminate those LRGs close to bright stars or nearby galaxies as described in \cite{2019ApJ...874..165Z}. 
The measured LRG SB profiles are fitted to Sersic models \citep{1963BAAA....6...41S} considering the stacked, extended DECam point spread function (PSF) averaged over DES Y3 observations. Figure~\ref{fig:profile} shows the stacked LRG $r$-band SB profile at $z\sim 0.25$ and the $i$-band profile at $z \sim 0.625$. The latter is corrected to $z\sim 0.25$ by the differences in luminosity distance moduli and angular-to-physical distance conversion. 

\section{Results}

We measure the LRG SB profiles with high S/N up to 200 kpc at both $z\sim 0.25$ and $z\sim 0.625$, and we do not notice significant evolution of the profiles between these redshifts, as they are consistent within $~2 \sigma$. We fit a single Sersic model, although the data only seem to follow this model within $\sim 60$ kpc, and less well at $z\sim 0.625$. 
At $z\sim 0.25$, the LRG SB profile is fitted by a Sersic model with index $2.59^{+0.05}_{-0.04}$ and effective radius $8.8^{+0.2}_{-0.1}$ kpc. At $z\sim 0.625$, the fitted Sersic model has an index of $2.75^{+0.03}_{-0.04}$ and an effective radius of $11.5^{+0.3}_{-0.5}$ kpc. The model fitting results are not sensitive to PSF convolution. Beyond 60 kpc, both the $z\sim 0.25$ and $z\sim 0.625$ LRGs show an excess of light above the Sersic models.

We derive the integrated LRG fluxes 
within 200 kpc, which are similar at $z\sim 0.25$ and $z\sim 0.625$ (corrected to the observer frame at $z\sim 0.25$) with values of $17.96 \pm 0.07$ mag and $17.93 \pm 0.03$ mag respectively. The integrated flux in the annulus between radius 60 and 200 kpc makes up 11.6\% of the total LRG flux within 200 kpc at $z\sim 0.25$ , or 9.7\% of the total flux at $z\sim 0.625$. In terms of the best-fitting Sersic models, the fluxes contained in the actual measurements but not in the models make up 10.0\% and 5.8\% of the total fluxes within 200 kpc at $z\sim 0.25$ and $z\sim 0.625$ respectively. We conclude that the extended light beyond 60 kpc, or the excess light not modeled by a single Sersic model, is not dominant but nevertheless still a nonnegligible LRG component.

A similar stacking analysis of LRGs has been reported in \cite{article} using SDSS data. They find that a Sersic model with $n=5.8$ and $r_e=13.6$ kpc fits the LRG profile well at $z\sim 0.34$ to $\sim 100$ kpc, which is different from our results. We suspect that the image processing methods, especially in terms of sky background estimations \citep{2017PASP..129k4502B, 2011AJ....142...31B}, may have played a role. We also consider how our LRG measurements differ from the intra-cluster light and central galaxy (ICL+CG) measurements at $z\sim 0.25$ in \cite{2019ApJ...874..165Z}. Unsurprisingly, the ICL+CG profile is brighter and its shape is more extented than the LRGs, as anticipated by the inside-out galaxy formation scenario in which CG starts out as a luminous compact galaxy and grows by merging in its peripheral regions \citep[e.g.,][]{2013MNRAS.435..901L, 2018MNRAS.479.1125R}. We have also analyzed the LRG $g-r$ ($z\sim  0.25$) and $r-i$ ($z\sim  0.625$) colors in our analysis, but unfortunately do not have enough signal-to-noise outside 20 kpc to draw robust conclusions.

This note is prepared under the DES publication guidelines\footnote{\url{ http://dbweb5.fnal.gov:8080/DESPub/app/PB/pub/pbpublished}}. A standard DES acknowledgment applies. 

\bibliography{reference}

\end{document}